\documentclass[12pt,preprint]{aastex}
\begin{document}

\title{The Jay Baum Rich telescope: a Centurion 28 at the Wise Observatory}

\author{Noah Brosch\altaffilmark{}, Shai Kaspi\altaffilmark{}}%, Dan Maoz\altaffilmark{} and Avishay Gal-Yam\altaffilmark{}}
\affil{The Wise Observatory and the Raymond and  Beverly Sackler School of Physics and
Astronomy, the Faculty of Exact
Sciences, \\ Tel Aviv University, Tel Aviv 69978, Israel}
\email{noah@wise.tau.ac.il, shai@wise.tau.ac.il}
\author{Saar Niv\altaffilmark{}}
\affil{Technion – Israel Institute of Technology, Haifa, 3200003 Israel}
\email{saarniv@gmail.com}
\author{Ilan Manulis\altaffilmark{}}
\affil{Martin S. Kraar observatory, Weizmann Institute of Science, Rehovot 76100, Israel}
\email{ilan.manulis@weizmann.ac.il}

%\maketitle
%\onehalfspacing
%\begin{singlespace}
%\end{singlespace}
%\twocolumn
%\columnsep 10pc

\begin{abstract}
%\begin{footnotesize}
We describe the third telescope of the Wise Observatory, a 0.70-m Centurion 28 (C28IL) installed in 2013 and named the Jay Baum Rich telescope to enhance significantly the wide-field imaging possibilities of the observatory. The telescope operates from a 5.5-m diameter dome and is equipped with a large-format red-sensitive CCD camera, offering a $\sim$one square degree imaged field sampled at 0".83 pixel$^{-1}$. The telescope was acquired to provide an alternative to the existing 1-m telescope for studies such as microlensing, photometry of transiting exo-planets, the follow-up of supernovae and other optical transients, and the detection of very low surface brightness extended features around galaxies.

The operation of the C28IL is robotic, requiring only the creation of a night observing plan that is loaded in the afternoon prior to the observations. %The observations were mostly performed remotely from the Tel Aviv campus or even from the observer's home, and now with robotic operation.
The entire facility was erected for a component and infrastructure cost of well under 300k\$ and a labor investment of about two person-year. %{\bf TBR We describe the two main projects to be undertaken with this new facility: the performance of a galaxy survey in H$\alpha$ and the monitoring of dense stellar fields to detect gravitational micro-lensing.}
 The successful implementation of the C28IL, at a reasonable cost, demonstrates the viability of small telescopes in an age of huge light-collectors.

\end{abstract}


\section{Introduction}
The Wise Observatory\footnote{http://wise-obs.tau.ac.il} started operations in 1971 and is the Tel-Aviv University (TAU) research laboratory in observational optical astronomy. It is located on the high plateau of the central part of the Negev desert (longitude 34$^{\circ}$.45'48" E, latitude 30$^{\circ}$35'45" N, altitude 875 m, in a time zone +2 hours relative to Universal Time). The site is 185-km south of Tel-Aviv and 86-km south of Beer Sheva, about 5-km west of the small town of Mitzpe Ramon.

\begin{figure}[t]
\centering{
 \includegraphics[width=16cm]{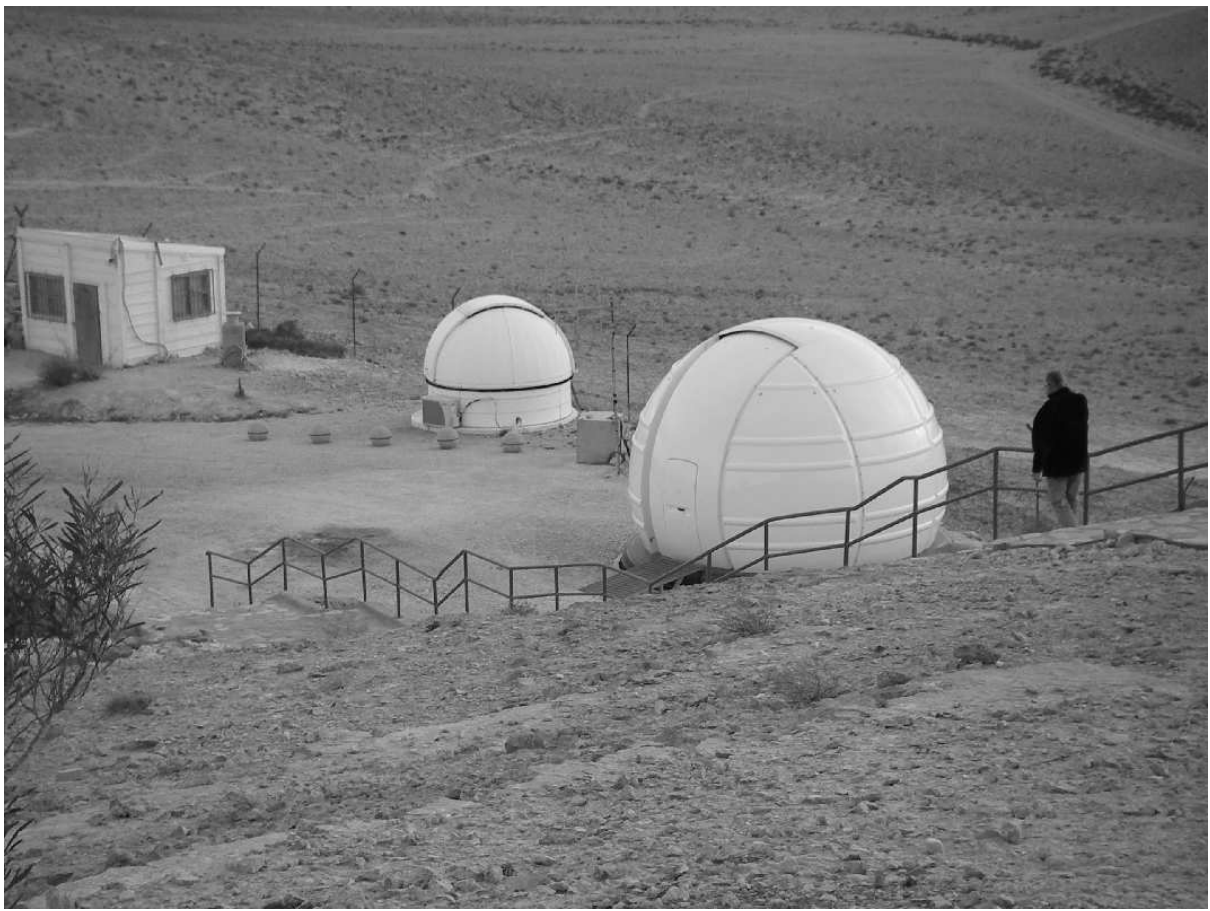}
 \caption{General view of the Centurion telescope domes at the Wise Observatory. The C18 dome is at the center of the image and the one for the C28 is at the right. The rectangular structure at the left is a storage shed.}
 \label{fig:WiseObs}}
\end{figure}

The WO is equipped since 1971 with a 40-inch telescope (T40), a widefield Ritchey-Chr\'{e}tien reflector mounted on a rigid, off-axis equatorial mount. %The optics are a Mount Wilson/Palomar
%Observatories design, consisting of a 40-inch diameter clear aperture f/4 primary mirror, a 20.1-inch diameter f/7 Ritchey-Chr\'{e}tien secondary mirror, and a quartz corrector lens located 4 inches below the surface of the primary mirror, offering a flat focal field of up to 2.5 degrees in diameter with a plate scale of 30 arcsec mm$^{-1}$. An f/13.5 Cassegrain secondary mirror is also available, but is presently not used.
 This telescope was originally a twin of the Las Campanas 1-m Swope telescope (Bowen and Vaughan 1973), but the two instruments diverged during the years due to individual modifications. The T40 telescope is operated through a control system located in the telescope room. % and, for most observations, it is used remotely.
%Since its inauguration, the observatory kept abreast of developments in the fields of detectors, data acquisition, and data analysis by upgrading its instrumentation and computer equipment. I
However, in many instances observations with the T40 can now be performed remotely via a secure internet link, freeing the observer from the necessity to travel to the observing site. The efficiency of modern detectors implies also that almost every photon collected by the telescope can be used for scientific analysis.

%The on-going modernization process allowed landmark studies to be performed and generations of students to be educated in the intricacies of astronomy and astrophysics. Some of these students are now staff members of the Physics and Astronomy Department at TAU, or at other academic institutions in Israel, or overseas. The one-meter telescope is over-subscribed, with applications for observing time exceeding by $\sim$50\% the number of available nights. This demonstrates the vitality of the observatory as a research and academic education facility, even though on a world scale the size of the telescope shrank from being a medium-sized one in the early-1970s to being a small telescope nowadays.

The Wise Observatory operates from a unique location, in a time zone between India and Greece and in a latitude range from the Caucasus to South Africa, where few modern observatories exist, and at a desert site offering a large fraction of clear nights. Thus, even though its 1-m telescope is now considered small, it is continuously producing invaluable data for the study of time-variable or time-critical phenomena, from data on Solar System objects and extrasolar planet searches to monitoring GRBs and microlensing events, to finding and following distant supernovae and ``weighing'' black holes in AGNs. %A cursory literature search shows that the Wise Observatory's T40 has one of the greatest scientific impacts among 1-m class telescopes.

%One research aspect that developed into a major WO activity branch is of time-series studies of astronomical phenomena. Other major projects include searches and follow-up observations of supernovae or extrasolar planets (using transits or lensing events), observations of novae and cataclysmic variables, studies of star-forming galaxies in a variety of environments, and studies of Near-Earth Objects (NEOs) and other asteroids.
 The studies done at the WO, mainly part of PhD projects, are observation-intensive and require guaranteed telescope access for a large number of nights and for a number of years. The oversubscription of the available nights on the T40, the need to follow-up possible discoveries %by small telescopes
  with long observing runs, and a desire
to provide a fallback capability in case of major technical problems with the T40, required therefore the expansion of the WO observational capabilities.
The T40 is now equipped with a high-quality Princeton Instruments camera, with a thinned, back-illuminated chip, but this instrument offers only a $\sim$13 arcmin field-of-view. To image wider fields the T40 can be equipped with the LAIWO instrument, a four-CCD mosaic of 4k$\times$4k pixels each. These chips, being thick and front-illuminated, offer a one-degree FOV but with QE$\leq$50\%. These reasons predicated the acquisition of a telescope and CCD camera capable of wide-field imaging with a higher QE than LAIWO.

Small automatic, or even robotic, instruments provide nowadays significant observing capability in many astronomy areas. Combining these small telescopes into larger networks can enhance their scientific impact.  Examples of such instruments operating in the previous decade are REM (Zerbi et al. 2001), WASP and super-WASP (e.g., Pollacco et al. 2006), HAT (Bakos et al. 2004), and ROTSE-III (e.g., Akerlof et al. 2003; Yost et al. 2006), and KELT (Pepper et al. 2007). Among the larger automatic instruments we mention the Liverpool Telescope (e.g., Steele et al. 2004) now part of the LCOGT network. For a collection of articles on robotic autonomous observatories see the proceedings of the III Workshop on Robotic Autonomous Observatories (Tello, Riva, Hiriart, Castro-Tirado 2014). Such robotic instruments enable the exploration of the last poorly studied field of astronomy: the temporal domain. However, a first step before developing networks of automatic telescopes is demonstrating their feasibility for a reasonable cost. % for a single, first instrument.
 This paper describes the second such experiment at the Wise Observatory in Israel.

We already described our first experiment, the installation and operation of a Centurion 18, which is an f/2.8 prime focus reflector (Brosch et al. 2008). During the past decade this telescope operated routinely in a semi-robotic mode using remote web access, requiring only the preparation of a list of observations (targets, filters, integration times, etc.) and the initiation of this list that is followed automatically throughout the night. The observer had only to open the dome and initialize the observing list, and close down at the end of the night while monitoring the weather and the instruments throughout the night. Since the beginning of 2015 the C18 telescope was switched to a fully-robotic mode of operation. A list of targets and observing constraints are uploaded during the day and the observations are carried out robotically throughout the night without requiring the intervention of the observer. %The same method has now been implemented for the C28IL, as will be described below.

%{\bf TBR We stress that, at present, there is no software link between the local weather station and the C18 operations, thus it is not possible to automatically close the dome and power down the telescope in case of inclement weather conditions. However, it is to be hoped that after the completion of the project described here, it will be possible to implement fully robotic operations of the C18.}

We describe below the newest addition to the WO observing arsenal, a Centurion 28 named the Jay Baum Rich telescope (hereafter C28IL) that provides enhanced imaging capabilities in a robotic mode of operation following the C18 precedent. The addition of this instrument to the WO telescopes was dictated by the necessity of a backup for the existing 1-m telescope in imaging studies. In particular, the projects for which the C20IL was needed are microlensing detection and follow-up, the following up of optical transients such as GRBs and supernovae, and the detection of low surface brightness extended features around relatively nearby galaxies.

The plan of the paper is as follows: we describe the telescope in Section~\ref{sec.scope}, the detector(s) in Section~\ref{sec.CCD}, the dome in Section~\ref{sec.dome}, the operating software and its integration with the hardware in Section~\ref{sec.sw}, we give an indication about the implementation cost in Section~\ref{sec.cost} and of the performance in Section~\ref{sec.perf}.%, and finally we present examples of science results obtained with the C28IL in Section~\ref{sec.results}.
 An image of the C28IL and the C18 telescope domes at the Wise Observatory is shown in Figure~\ref{fig:WiseObs}.

\section {The telescope}
\label{sec.scope}

Following the acquisition of the C18 telescope manufactured by AstroWorks, USA, and its delivery
towards the end of 2003, this telescope operates since 2005 in its permanent dome. The C18 has
a prime-focus design with an 18-inch (0.46-m) hyperbolic primary (lightweighted) mirror figured to provide a wide-field f/2.8 focus using a Ross doublet corrector lens with SF2 and BK7 glasses. The focal plane is maintained at the proper distance from the primary mirror by a carbon-reinforced epoxy plastic (CREP) truss tube structure that allows the routing of various wires and tubes to the focal plane through the CREP tube structure, providing a neat construction and much less vignetting than otherwise. To economize, we operate the C18 from a 10-foot ProDome; this small enclosure is sufficient for the telescope and its computer but makes maintenance work very difficult. For the new telescope we selected a more spacious housing.

We learned that the same manufacturer produced a small series of scaled-up versions of the C18, and that the first one of this model was acquired and was operating in the US. In fact, some early results from that telescope were published in \textsl{Nature} (Rich et al. 2012). Such a telescope, when equipped with a good CCD, could provide wide-field imaging and replace or supplement the T40 while providing high-efficiency imaging. This instrument, a 28-inch, operates in the US at the Polaris Observatory Association, Frazier Park, California. Hereafter, this telescope will be referred to as the C28US.

The C28IL is essentially a scaled-up version of the C18 described in Brosch et al. (2008) operating at f/3.2. Its optical assembly is supported by a fork mount similar to that of the C18 but larger and sturdier, with the right ascension (RA) and declination (DEC) aluminum disk drives being of pressure-roller types. One difference from the C18 is that the C28IL is equipped with tensioned phosphoric bronze rings over the aluminum disks, and the stainless steel rollers press on these rings to provide stiffer and sturdier drives. The steel rollers are rotated by stepping motors and both axes are equipped with 13-bit absolute optical sensors yielding a ``cold start'' pointing accuracy of $\sim$3 arcmin. The sensors, the stepping motors, and the FS-2 controller (from Astro Electronic in Germany\footnote{http://www.astro-electronic.de/}) ensure that the motions of the telescope can be controlled fairly accurately by a computer. In addition, and as a safety measure, optical gates are installed on both axes to prevent uncontrolled driving of the telescope into the dome floor. %, and to define a ``zero'' position to which the telescope is initialized prior to observations.

The primary mirror of the C28IL is a lightweighted ribbed structure. It was manufactured by spin casting, followed by polishing and coating. The spin-cast fabrication left a few air bubbles in the softened glass that are visible after polishing and coating as small ``craters'' in the aluminized surface; they are small and few, and do not affect significantly the scientific operation. The mirror uses a ``protected aluminum'' coating, with a somewhat higher reflectivity in the spectral range of interest than bare aluminum.

The focal plane assembly is shielded from stray light by a conical baffle and
 permits fine focusing using a computer-controlled focuser (RoboFocus) moving the doublet corrector lenses with a stepper motor. %The C28IL was originally supplied with a primary mirror metal cover that had to be removed manually. Since then, an electrically-operated, remotely-commanded mirror cover was installed {\bf TBR}.
 We decided to equip the C28IL with a modern, large-format, CCD camera described below. Unlike the SBIG STL-6303 CCD used in the past few years with the C18 telescope, which has a guiding CCD adjacent and in the same enclosure as the science CCD, the C28IL camera has only the science CCD at the focal plane with no space in its box to add a guiding camera. For this reason, the C28IL was equipped with a bore-sighted refractor telescope which serves as offset guider.  The guider telescope is an Orion AstroView 120ST telescope, with an aperture of 120-mm and a focal length of 600-mm (f/5). %an  for guiding.

The guider camera at the focus of the Orion telescope is a miniature SBIG STi CCD camera equipped with a Kodak KAI-340 chip with 648$\times$484 anti-blooming pixels each 7.4 $\mu$m square, yielding a field of view of 27'.2$\times$20'.4 with $\sim$2.5-arcsec pixels. This camera is operated uncooled and is very suitable for guiding. Given the relatively short focal and physical length of the guiding telescope, and its mounting close to the telescope center-of-mass near the mirror, we do not experience significant differential flexing between the two optical assemblies.

\section {The science CCD}
\label{sec.CCD}

The science CCD is mounted behind the doublet corrector lens and the focusing is achieved by moving the lens with a total travel of $\sim$seven mm covered by $\sim$2200 steps of the focus motor. In practice, even though the beam from the telescope is strongly converging, the focus is fairly stable throughout the night despite ambient temperature excursions of $\sim$10$^{\circ}$C or more. In any case, refocusing is very easy using the robotic focuser.

The camera is a Finger Lakes Instruments (FLI) ProLine PL16801 product equipped with a thick, front-illuminated Kodak KAF-16801 chip. The chip has 4096$\times$4096 pixel format with 9 $\mu$m square pixels,  translating at the plate scale of the C28IL into $\sim$0.83-arcsec pixel$^{-1}$, covering a field-of-view of $\sim$57$\times$57-arcmin$^2$. The quantum efficiently (QE) of this chip peaks at $\lambda\simeq$661 nm with QE=67\% and the QE is higher than 50\% from 520 to 730 nm. The camera includes a built-in Uniblitz 65-mm optical shutter with an opening time of 22 msec.

The FLI camera is thermoelectrically-cooled (TEC), with water circulation cooling the hot ends of the TEC cooler. This avoids discharging hot air at the sky end of the telescope from the TEC, improving the telescope seeing. Depending on the ambient conditions, the chip temperature ranges between  --25C (summer) to --35C (winter), which changes negligibly the dark current. The cooling water is pumped up to the prime focus and the FLI camera from a container resting on the dome floor via 5-mm diameter flexible tubes inserted in the CREP tubes that make the telescope truss assembly, and returns to this container through similar tubing.

FLI has installed as a default option for this chip a ``preflashing'' option. This, because the KAF-16801 has a known residual bulk image (RBI) whereby traces of previously exposed fields can appear in a new exposure of a different field. The RBI is caused by electrons drained by the substrate with some captured in charge traps, which appeared in the CCD substrate during the CCD manufacturing. The charge accumulated in these traps slowly disappears, but some of these electrons move back to the image pixels and create ``ghost'' images of previously-imaged fields at locations of bright stars.

The preflashing that solves this issue is implemented at the start of a new exposure: the chip is illuminated with light from surrounding near-IR LEDs and it is read out quickly twice before the ``real'' exposure starts. This ensures that the charge traps responsible for the latent image are fully filled when the new image starts.

The KAF-16801 was selected to exploit its significantly higher QE at $\lambda\geq$660 nm than the similar and more popular chip KAF-16803, at the expense of slightly reduced QE at $\lambda\leq$400 nm. This, because of the requirements of the two major research programs to be undertaken with the C28IL: microlensing toward the Galactic Center and deep imaging of nearby galaxies.

The camera is equipped with a five-position computer-controlled filter wheel (FLI CFW-9-5) that interfaces smoothly with the camera, is computer-controlled, and accepts filters up to 65$\times$65-mm and up to 5-mm thickness. Special adapters allow the use of smaller filters, at the expense of additional vignetting.

\section{The dome and ancillary equipment}
\label{sec.dome}

Given the high degree of automation required, we chose a dome permitting daytime maintenance operations with humans inside, while allowing night-time unrestricted access to the sky for the C28IL. From among the off-the-shelf domes we chose a product of ScopeDome\footnote{http://www.scopedome.com}, a Polish astronomical dome manufacturer. The ScopeDome 5.5-m diameter dome is a fiberglass product manufactured and shipped in segments, and assembled on-site. The dome is in the shape of a 3/4 sphere and is equipped with an electrically-operated over-the-top 1.7-m wide shutter.

The dome azimuth motion is effected by a 5-m diameter toothed ring that is engaged by the drive motor. The entire fiberglass dome is mounted on this base ring and moves as a single entity. The dome shutter moves up and down on rollers travelling in special channels built into the dome surface and its motion is achieved by a second motor mounted on the dome, which pulls the shutter through a slotted rail internally mounted on the shutter.

The  version of the ScopeDome used for the C28IL is equipped with a control system (hereafter CS) that offers manual control via push-buttons as well as the computer-controlled operation of the dome, since it and its driving software are ASCOM\footnote{http://ascom-standards.org/}-compatible. The CS interfaces with the dome azimuth motor, the shutter motor, and the various limit switches and dome encoders, connecting them with the control computer. The CS has two parts: one is stationary and is mounted near the telescope and the other is mobile and is mounted on the dome. The stationary part deals with the dome azimuth motion while the mobile part controls the shutter motion. The two parts communicate via a wireless link. In addition, the CS has internal and external temperature and humidity sensors as well as a barometric pressure sensor, all mounted on the mobile part of the CS. The stationary part of the CS is connected to the control computer via a USB card.

Since the FLI CCD camera that is used for science observations has no provision for a parallel small CCD in the same housing, which could be used for guiding, and there is no clearance in the optical path to add an optical pickup system for a small guiding CCD, we decided to add a parallel telescope equipped with a CCD that would provide the guiding signals. The telescope is an Orion 5-inch refractor operating at f/5 and imaging on a miniature CCD camera (SBIG monochrome STi). The 7.4 $\mu$m pixels translate to a plate scale of $\sim$2".5 pixel$^{-1}$ and the field covered by this configuration is $\sim 27 \times 20$ arcmin$^2$. This offset guider is attached to the telescope via two sturdy rings and we did not detect any differential flexing between the main telescope and the offset guider at all orientations.

%The ScopeDome software can access the output of %the observatory's weather station mounted on the T40 main building. In addition, we provided the C28IL dome with
 The C28IL has its own weather station located some 10-m away from the dome, consisting of a Boltwood Cloud Sensor II, which provides wind speed, % and direction,
  humidity and temperature, a calculated dew point, and an indication about the cloudiness from a special cloud sensor that measures the amount of radiation in the 8 to 14 micron infrared band received from the sky and from the ground. A large difference indicates clear skies, whereas a small difference indicates dense, low-level clouds. The 5.5-m dome is equipped with an internal web camera and with a commandable light source that allows remote viewing of the telescope and of part of the dome and its shutter to check their position and status (shutter open/closed). %, and a "Robo reboot" device for DDW. The latter was installed to allow the remote initialization of the dome functions in case of a power failure.
 All systems of the C28IL observatory (telescope,  CCDs and operating computer) are connected to the mains power supply through a "smart" UPS that buffers mains power outages and allow the continuing operation or the parking of the telescope in case of an extended power outage.

\begin{figure}[t]
\centering{
 \includegraphics[width=16cm]{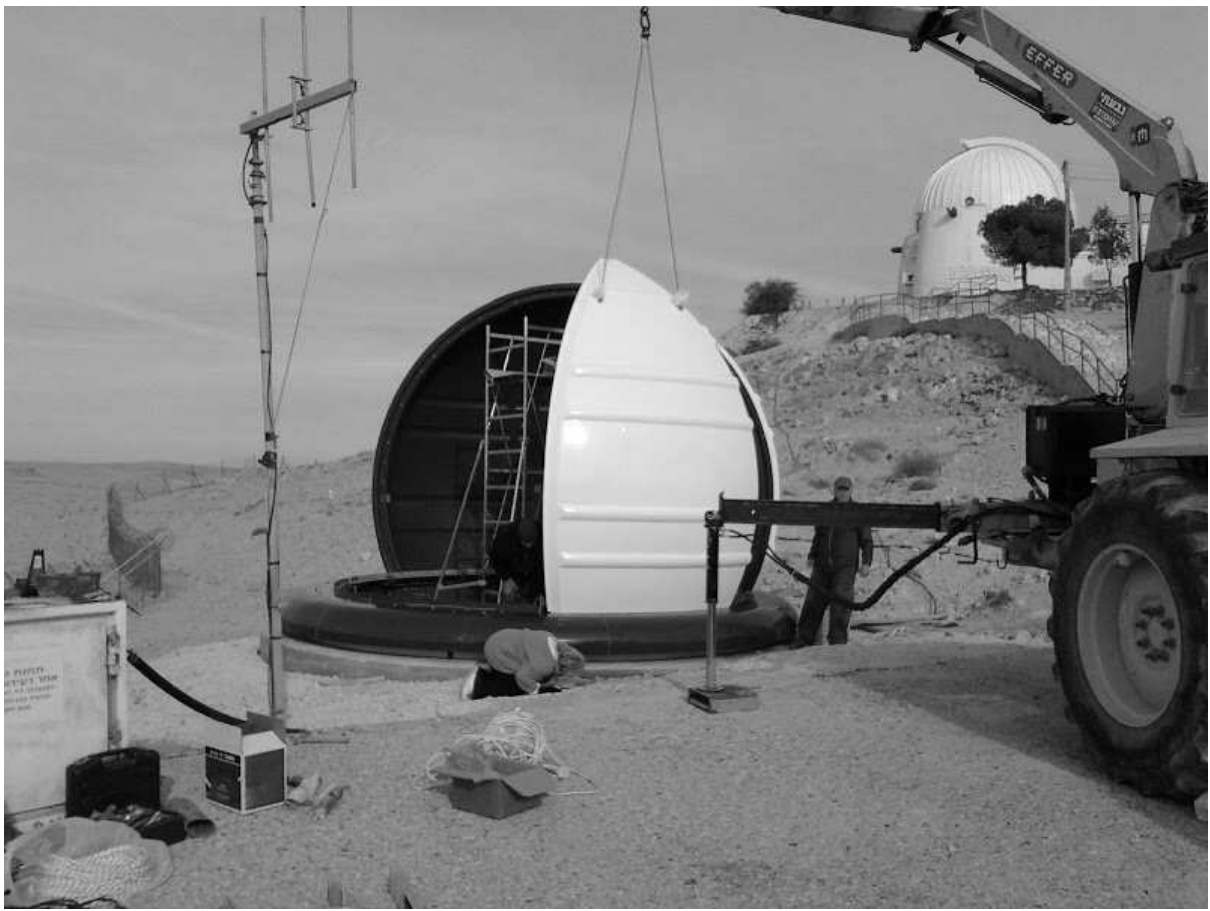}
 \caption{Construction of the C28IL dome. The concrete base is visible under the skirt of the partly-completed dome, while the crane is supporting one of the dome segments prior to connecting it to the rest of the dome. The antenna seen in the foreground was moved since it could, in some situations, block the telescope view.}
 \label{fig:Dome}}
\end{figure}

The dome and telescope are mounted on a circular reinforced concrete slab with a 6-m diameter and 0.25-m thickness, with its outermost 50-cm edges deeper by 0.3-m. The concrete was poured directly on the bedrock, which forms the surface ground layer at the WO site. The telescope is mounted slightly off-center in the dome, on a segment of this slab that is thicker than the rest of the floor (0.8-m) and is mechanically isolated from the rest of the slab to prevent the transmission of vibrations to the telescope when the dome moves. The dome, as it was when under construction, and the concrete slab on which it rests, are shown in Figure~\ref{fig:Dome}.

\begin{figure}[t]
\centering{
 \includegraphics[width=16cm]{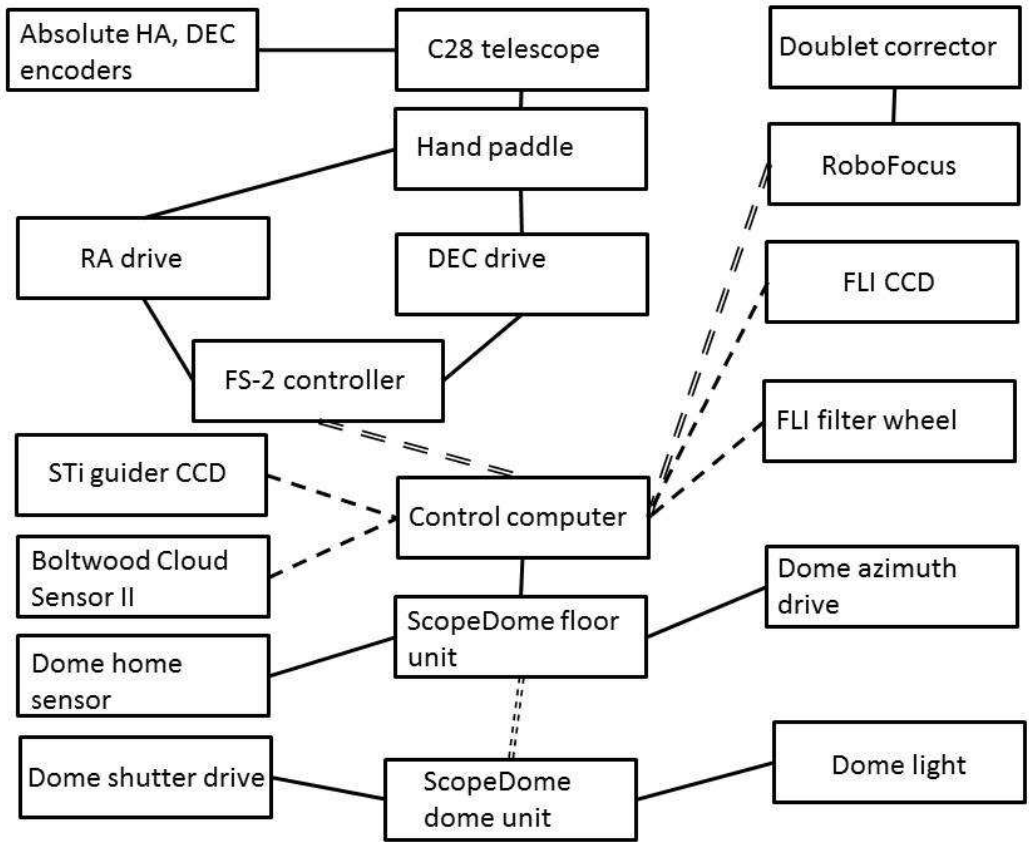}
 \caption{Automation scheme of the C28IL telescope (hardware and connections). The various connections are coded as continuous line=electrical hardwired connection, dashed=USB, double-dashed=serial link, double-dotted=radio link. }
 \label{fig:AutomationHW-conn}}
\end{figure}

With the FLI CCD described here at the prime focus, there is a $\sim$10-cm clearance space between the back end of the CCD at the prime focus and the lowest parts at the top of dome. There are no limitations to pointing the telescope anywhere on the celestial sphere above the lowest part of the shutter assembly (altitude $\sim20^{\circ}$ above the local horizon), but there is partial blockage by the dome shutter for objects located very close to the zenith. This implies that objects passing within $\sim10^{\circ}$ of the zenith of the C28IL must be observed while they are about one hour away from meridian passage. In addition, while the North Celestial Pole is observable by the C28, there is slight terrain blockage by the T40 dome and building to the North with the telescope below the polar axis. The various hardware components of the C28IL telescope complex automation scheme and their inter-connections are shown in Figure~\ref{fig:AutomationHW-conn}.

%The status of the telescope and dome can be checked by the remote observer using a webcam equipped with a fisheye lens, while the interior of the dome is illuminated by a remotely-controlled halogen lamp.

\section {The operating software}
\label{sec.sw}

\begin{figure}[t]
\centering{
 \includegraphics[width=14cm]{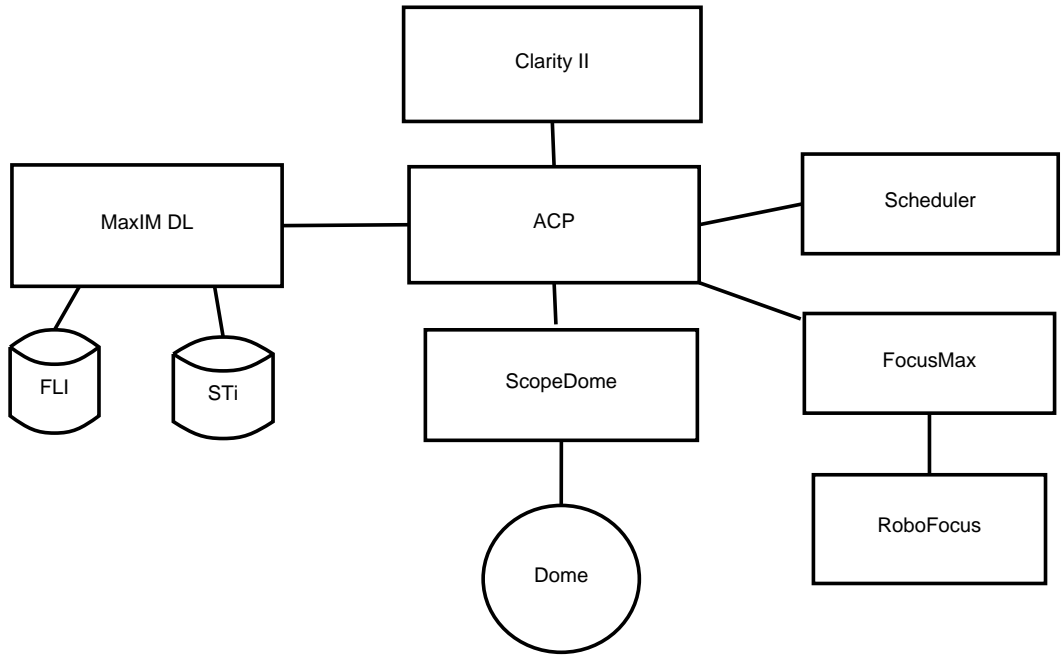}
 \caption{Control scheme for the C28IL telescope (software and hardware links).}
 \label{fig:SW-conn}}
\end{figure}

As for the C18, we decided that all the software would be tailored into a suite of operating programs that would conform to the ASCOM standards. This is in contrast with other similar%, but significantly more expensive,
 small robotic observatories (e.g., Akerlof et al. 2003), which chose various flavors of LINUX and special software. Our choice saved the cost in %money and
 time of developing specialized software by using off-the-shelf products. It also facilitated a standard interface to a range of astronomy equipment including the dome, the C28IL mount, the focuser and the camera, all operating on a single computer in a Microsoft Windows/ASCOM environment. The general software scheme is described in Figure~\ref{fig:SW-conn}.

The general operation of the C28IL and its associated devices is run by the Scheduler, which is part of the Astronomer's Control Program (ACP), a product of DC-3 Dreams\footnote{http://acpx.dc3.com/}. The Scheduler accepts inputs as a list of observing requests (objects with coordinates, telescope and CCD setups, exposure times, etc.) and a set of constraints to perform the observations (altitude of the object, seeing, sky brightness, etc.). With these, the Scheduler decides automatically when and what to observe. The Scheduler provides also set interrupts that allow quick and efficient re-scheduling in case of urgent time allocations to observe transient events (e.g., for GRB, microlensing and supernova follow-ups).

%{\bf TBR}
The ACP software controls the telescope motion and pointing, and can change automatically between different sky fields according to instructions from the Scheduler. ACP also solves astrometricaly the images collected by the CCD via the PinPoint7 program. %, and improves automatically the pointing of the telescopes using these solutions.
 ACP communicates with and controls the operating software of the dome, interfacing with ScopeDome, enabling one to open and close the dome shutter, and commanding the dome to follow the telescope or to go to its "home" position.

 In addition, the ACP software is a gateway to the MaximDL program that operates the CCD. Different types of exposures, CCD preflashing, guiding and cooling of the CCD can be commanded manually using MaximDL, but in most cases we use the ACP code envelope through the Scheduler to operate the entire system. The focusing is also done robotically using the freeware FocusMax, a software package that operates the Robo-focus and searches automatically for the best FWHM of a selected star.

%{\bf TBR} To enable power shut-down by remote users we connected the telescope, dome, CCD, focuser and the telescope's cover to a relay box that is connected to the six serial ports on the computer. Using a self-written code (C18 control) the remote user can enable or disable the electrical power supply to the different components of the system and open or close the telescope cover. This guarantees the safety of the equipment during daytime and enables the astronomer to fully operate the system from a remote location using any VNC viewer software. %Figure 5 exhibits the different programs making up the software environment, their connections and their hierarchy.

The Clarity II program runs the Boltwood Cloud Sensor II weather station providing current and historical weather reports, as well as threshold values of wind speed, relative humidity, and cloudiness. These are used by the ACP with Scheduler to close down the C28IL in case of inclement weather.

The above-described system allows the C28IL to operate robotically. The Scheduler accepts the list of targets and of observational constraints during the day. It then powers-on all the systems in the late-afternoon, commands the opening of the dome after sunset, initializes the telescope and CCD cameras, and takes calibration images (bias, dark and sky flat fields). After sunset the Scheduler also focuses the telescope via FocusMax, and carries out the planned observations. In the morning the Scheduler parks the telescope, closes all systems and the dome, and powers everything off. In case of inclement weather conditions (wind, humidity, or cloudiness limits exceeded), the Scheduler closes the dome, and opens it again only after the weather conditions become safe for observations, checking for this every ten minutes.

\section {Cost}
\label{sec.cost}

The affordability of a robotic telescope is an important consideration for many observatories. We did not have sufficient funds to cover the high cost of an off-the-shelf robotic telescope with similar capabilities to those of the installation described here.

The C28IL telescope, including all the electronic add-ons and the software, added up to slightly more than 120 k\$. The ScopeDome cost was 25 kEuro, including the full automation package and wiring harnesses. The FLI ProLine PL16801 and its filter wheel were some 27 k\$. To these we must add taxes and local expenses such as ground levelling and filling, casting of the telescope pier and of the concrete dome base, the air conditioner, the internal electrical installations, and the operating computer. A total of about 250 k\$ for the entire installation would not be too excessive.

There were also significant in-house contributions; the dome was received in segments that were erected and bolted on the concrete slab by WO staff Shai Kaspi, Ezra Mash'al, Sammy Ben-Guigui and WO's friends Assaf Berwald, Ilan Manulis and Evgeny Gorbikov together with NB; the installation, tuning-up and interfacing of the various software components were done mainly by IM and SK; the automation and robotization were done primarily by SN, the polar alignment of the C18 was done by IM and AB. The crane used to assemble the dome and its operator were kindly loaned free-of-charge by the local municipality. %AB also developed the electrically-operated primary mirror cover for the C18, on which the C28IL cover was patterned.
 All the human contributions add up to about two person-years of work by very experienced personnel.

\section {Performance}
\label{sec.perf}

\begin{figure}[t]
\centering{
 \includegraphics[width=14cm]{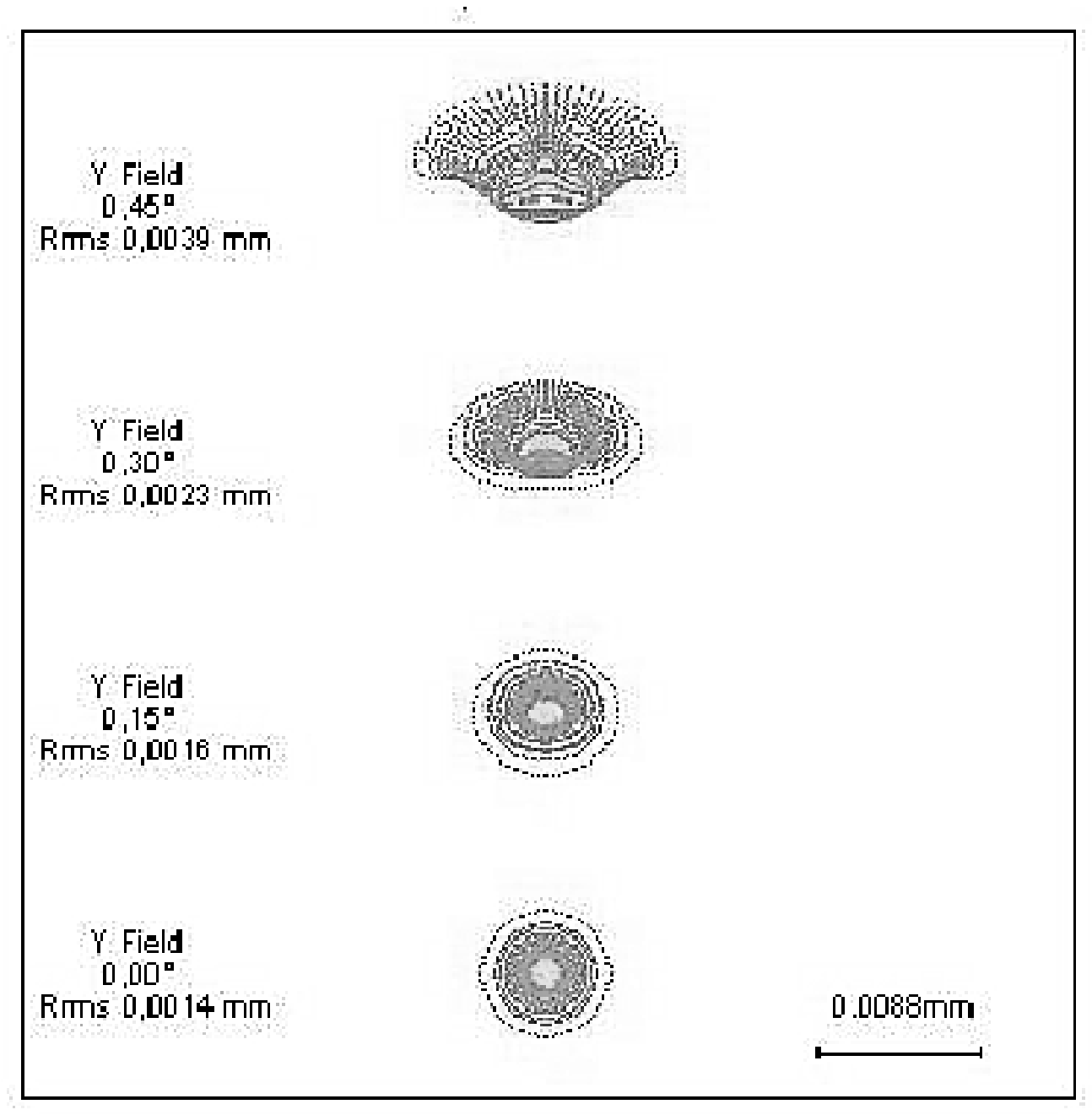}
 \caption{Polychromatic spot diagram for the C28IL.}
 \label{fig:poly-spot}}
\end{figure}

The optical performance of the C28IL was simulated with ray tracing. A polychromatic spot diagram predicted for the telescope is shown in Figure~\ref{fig:poly-spot}. The bar at the bottom-right of the figure is approximately the size of one FLI CCD pixel. The diagram shows that, theoretically, the C28IL produces images smaller than one pixel in the inner 36 arcmin of the frame, and that even at the edge of the field the images are about two pixels wide. The telescope image quality and the FLI pixel size match nicely the typical seeing at the Wise Observatory, 2-2.5 arcsec=2.4-3 pixels FWHM.
The robotic operation mode of the C28IL makes it an easy telescope to use. %, requiring the astronomer only to monitor the weather conditions when these are likely to change and this only .
Overall, the C28IL's performance and output are satisfying.

The telescope slewing time is about one deg s$^{-1}$ on both axes and the nominal settling-down time is $\sim$3 s. The dome rotates at a rate of two deg s$^{-1}$ in each direction. Opening or closing the dome shutter requires about one minute, with the dome parked at its ``home'' location. %Opening or closing the electrical telescope cover requires 20 s.

%{\bf TBR} The auto-guiding system, which uses a smaller CCD in the same SBIG STL 6303E camera head, maintains round stellar images even for the longest exposures that are limited by the sky background, although the telescope is not perfectly aligned to the North. Without the guider, one can expect round star images only for exposure times of 90 sec or shorter. %In some cases, telescope shake is experienced due to wind blows. Since the wind at the WO usually blows from the North- West, and usually slows down a few hours into the night, selecting targets away from this direction for the first half of the night decreases the number of smeared images to a minimum.

The initial pointing errors of the C28IL are of order 10 seconds of time in RA and 30 to 60 arcsec in DEC. These are reduced by an order of magnitude following the first image of the night taken by the CCD camera, which is solved astrometrically and the telescope pointing is adjusted. Astrometrically solving all the images using the PinPoint7 engine, operated automatically by  ACP immediately following the image readout, re-points the telescope to a more accurate position for the following images of the same field.

Since the CCD cooling is thermoelectric with water assistance, with the cooling water at ambient temperature, the CCD temperature depends somewhat on the weather, with a chip temperature of --25C in summer nights with ambient temperature above +25C and the cooler running at less than 90\% efficiency. %. The usual chip temperatures run between --15$^{\circ}$C in the summer nights
 However, the dark current hardly changes with ambient temperature for CCD temperatures below --20C. %s of +30$^{\circ}$C to --30$^{\circ}$C in the winter (with +5$^{\circ}$C ambient).
  All images are acquired with the CCD cooling-power at less then 100\% capacity, assuring a steady chip temperature throughout the nightly observation.

The full image read-out of the FLI CCD, using MaximDL, with no on-chip binning and reading out the entire chip at 8 Mpixel s$^{-1}$, requires $\sim$15 s. In regular observations an additional 10 s interval is required for the astrometric solution of the image and another 5 s to write the image on the local computer. A five-sec delay is required by the CCD before it continues to the next image for guider activation, in case the exposures are not dithered automatically. The auto-dithering requires the re-acquisition of the guide star and adds $\sim$30 sec to the overhead. Since the performance of the astrometric solution and the activation of the auto-guiding operation are user-selectable,
the off-target time between sequential images is 15--40 s, yielding a duty fraction of 83--92\%  on-sky time for typical 300 s exposures. Invoking the preflash option adds $\sim$30 s, because of the double readout at 8 Mpixel s$^{-1}$, thus reducing further the on-sky duty fraction.

\begin{figure}[t]
\centering{
 \includegraphics[width=14cm]{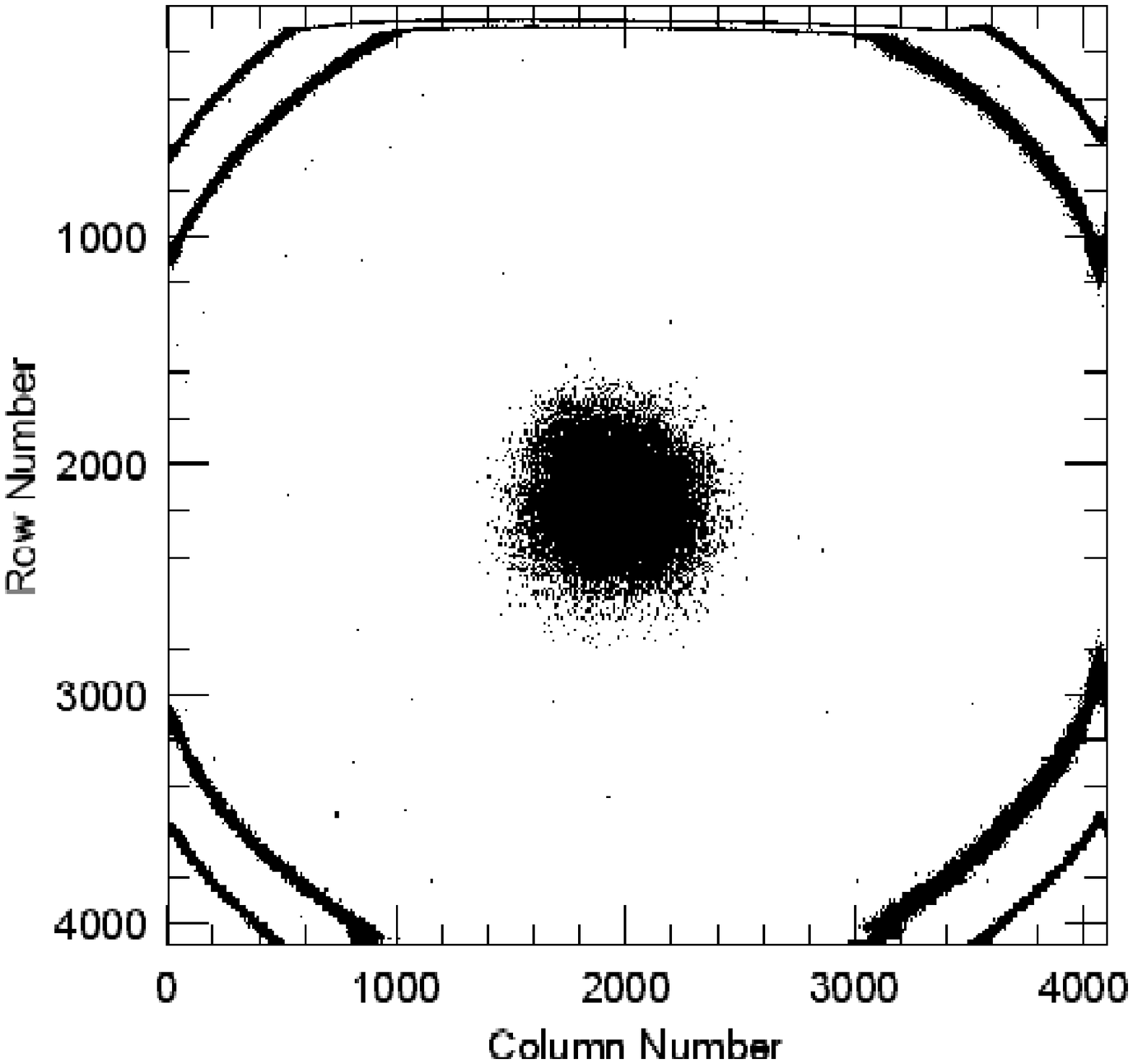}
 \caption{Normalized sky flat field in the L-band. The contour levels are 0.9, 0.95, and 1.03.}
 \label{fig:L_FF}}
\end{figure}

The CCD bias values are about 1000 counts. %, a value which is somewhat cooling-dependent. %The bias values changed from an original ¡«110 to the present value after one year of use and a rebuild of the software following a disk crash.
 The CCD flat field (FF) shows slight vignetting at the corners, as demonstrated
in Figure~\ref{fig:L_FF}. The image has been normalized to a global value of 1.0 and shows that the very corners of the image are at a level of 0.9 and lower. %This probably reflects some additional vignetting in the telescope on top of the prime focus baffle, possibly caused by the asymmetric blockage by the CCD and its filter wheel.% , and by the pick-off prism edge for the TC-237H tracking CCD, mounted next to the science CCD, which is used for guiding.
 Twilight flats give the best results, provided the CCD cooling is started at least 30 min before obtaining the flat field images. The corner vignetting of the image when using the 65$\times$65-mm filters is at most $\sim$82\%.

A preliminary transformation from the luminance L-band  to the SDSS-r band, %based on measurements of the Landolt sequence PG0918+029,
 is
\begin{equation}
r\simeq22.68(\pm0.11)+0.97(\pm0.02)\times[-2.5\times log(L)]
\end{equation}
where $L$ is the net counts-per-second from the source through the L-band filter. A small color term dependence on ($g-r$) has not yet been fully determined.

On a dark night with photometric conditions the sky background is $\sim$10-–15 counts pixel$^{-1}$ s$^{-1}$. Using R-band magnitudes of standard stars (Landolt 1992) the sky as measured on the C28IL images in the ``wide-R'' Luminance band is about 20.4 mag arcsec$^{-1}$; stars of 19.5 magnitudes are detectable with S/N$\simeq$25 with a 120-s exposure. This fits well previous measurements of the sky brightness at the Wise Observatory (R$\simeq$21.2 mag arcsec$^{-1}$ in 1989; Brosch 1992), accounting for the slight site deterioration with time due to the addition of new ambient light sources and for the wider spectral bandpass of the L filter.

\begin{figure}[t]
\centering{
 \includegraphics[width=14cm]{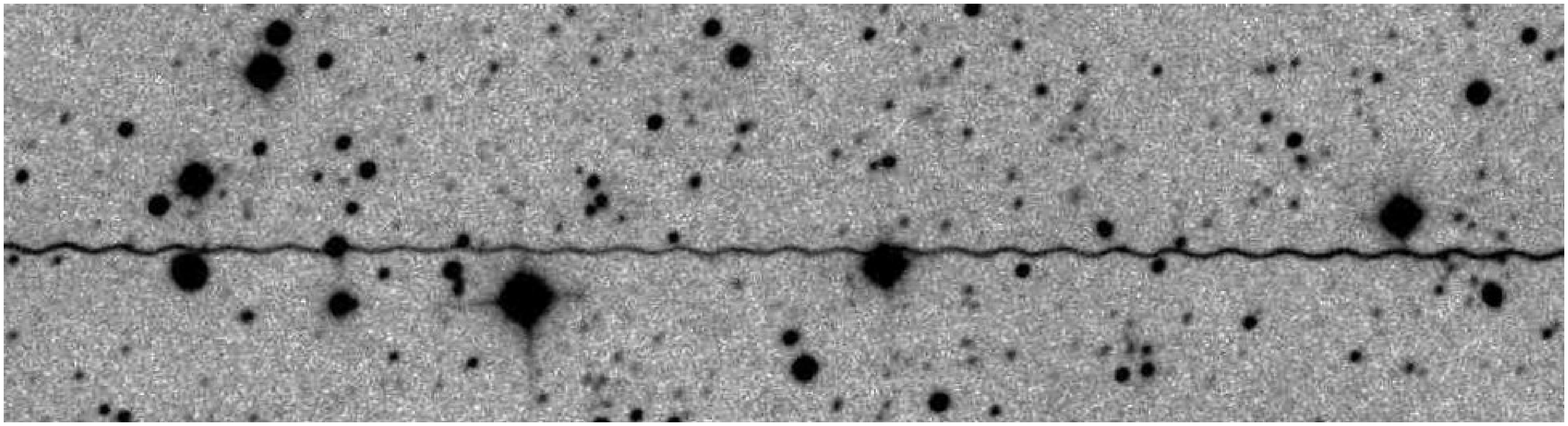}
 \caption{Enlarged segment of image containing UGC 9169 (which is out of the image), and a satellite track that demonstrates the very low-level telescope oscillatory behavior. The peak-to-peak oscillations should be compared to the visible stellar image FWHM to evaluate the influence on the seeing.}
 \label{fig:U9169_oscil_track}}
\end{figure}

We normally operate the C28 with guided exposures of 180 to 900 s duration. The guiding CCD (STi) is read out every five seconds and the rms guiding errors are less than 0.2 STi pixels, i.e. about 0.5 arcsec. We discovered, by analyzing satellite tracks taken early in the night or before the morning twilight that the telescope and camera record small periodic or quasi-periodic oscillations. An example of such oscillations, recorded while imaging the galaxies NGC 5022 and NGC 5018, is shown in Figure~\ref{fig:U9169_oscil_track}. It clear that this could not be the signature of a tumbling or spinning low Earth orbit satellite, thus this could be caused either by a periodic drive error (on both RA and DEC, or by some unknown induced vibration.

The frequency of the oscillations is about 80 Hz and the peak-to-peak amplitude seems to be slightly smaller than the full width at half-maximum of the stellar images (in this image $\sim$4".3). However, it is not clear that these oscillations are always present, at every hour angle and declination, and that they always have the same amplitude. If they do, they could be responsible for the slight deterioration of the stellar PSF, on top of the site seeing. The origin of the oscillations has not yet been identified.

Despite all these issues, the telescope is capable of producing good science. We show below an example, obtained during the telescope commissioning, when the Coma Cluster of galaxies was imaged, 28 images were collected through a Luminance filter each exposed for 300 s. The images were debiased, dark-subtracted, flat-fielded with twilight flats, registered, and co-added. The image shown here is a small section of the entire image displayed to show the tidal stream discovered by Gregg \& West (1998).

\begin{figure}[t]
\centering{
 \includegraphics[width=14cm]{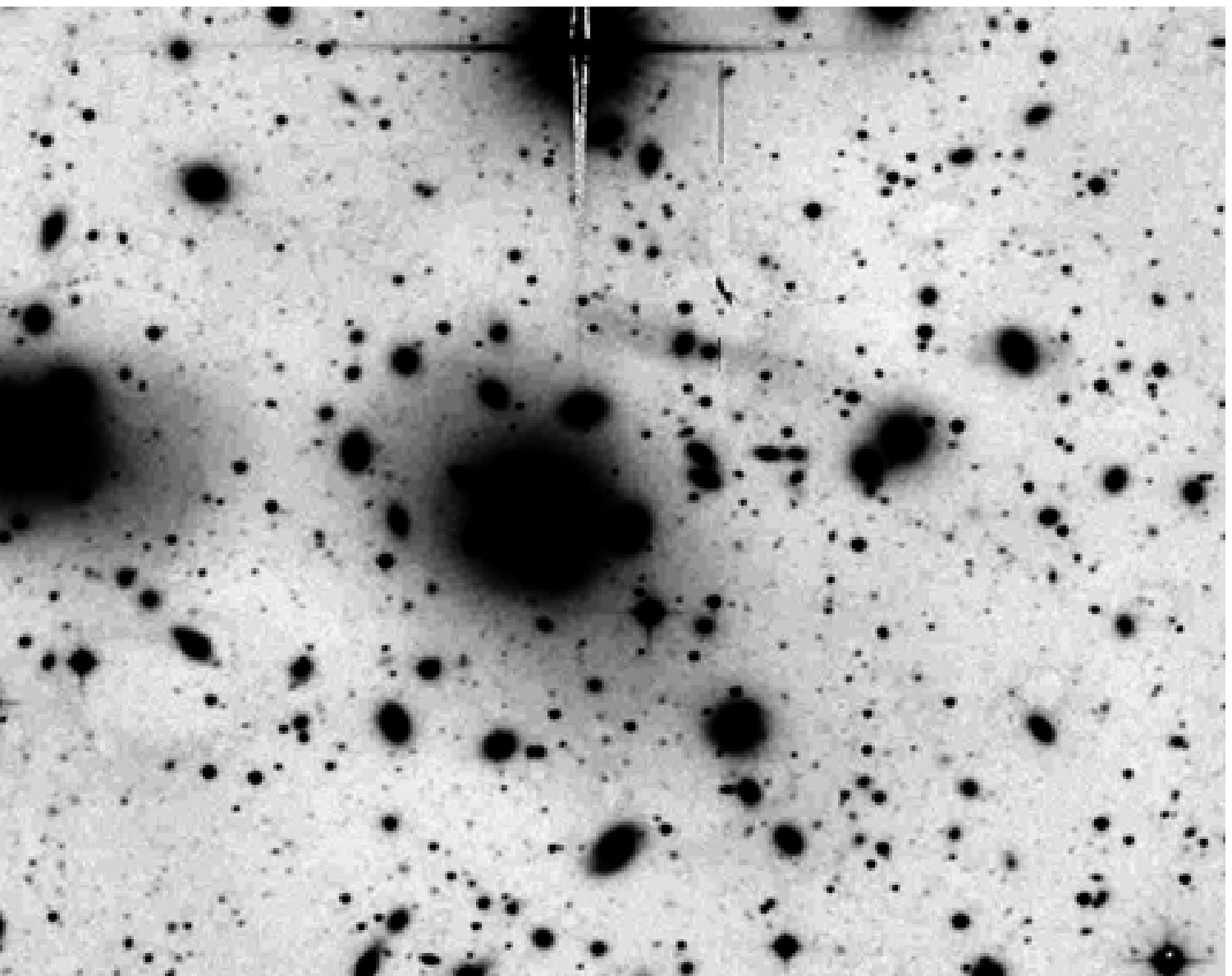}
 \caption{Enlarged segment of a deep C28IL image containing NGC 4876 (center) and the tidal stream north (up) of the galaxy.}
 \label{fig:N5022_oscil_track}}
\end{figure}

\begin{figure}[t]
\centering{
 \includegraphics[width=14cm]{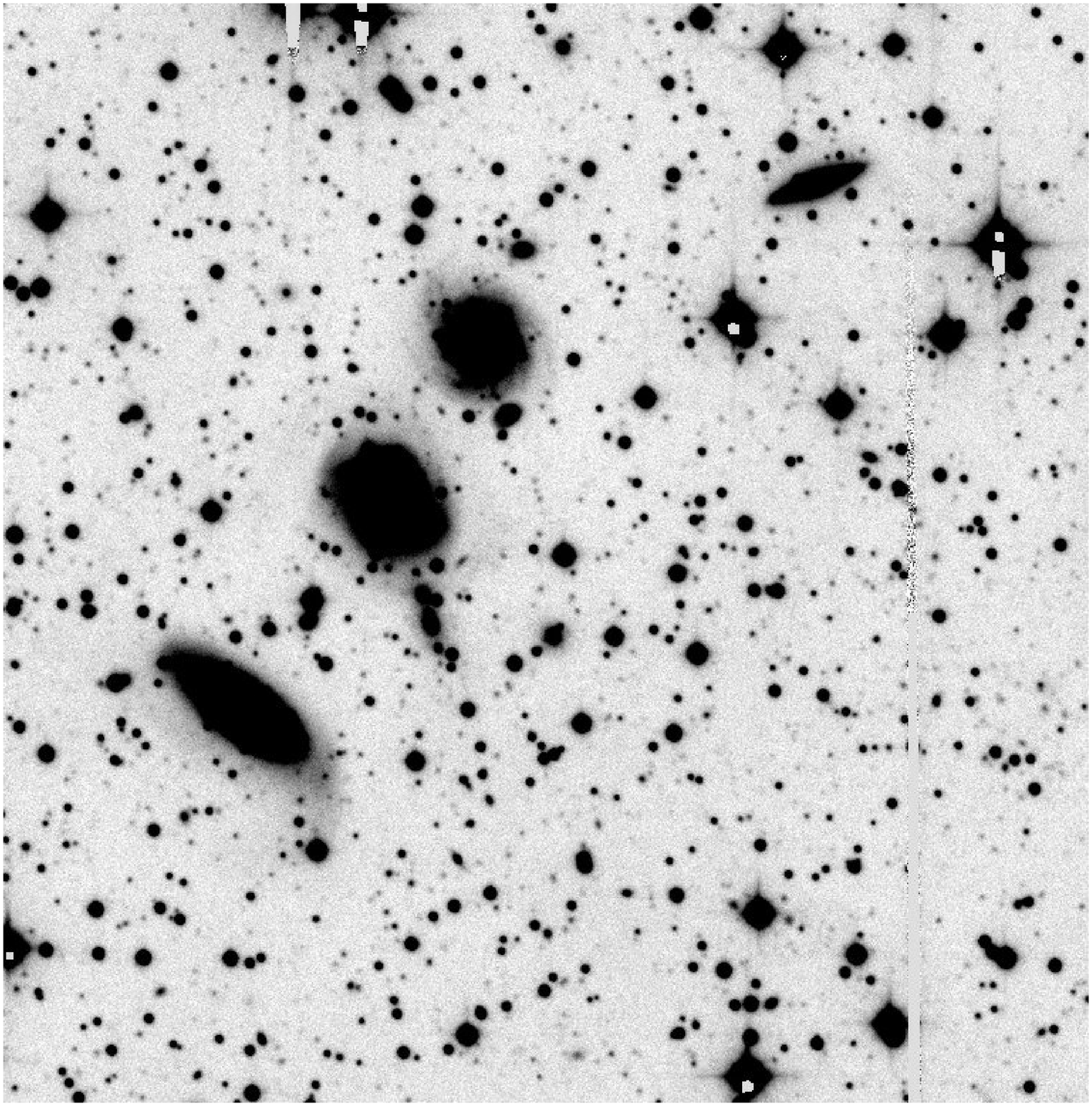}
 \caption{The Hickson Compact Group 88. At least two galaxies shows sign of interaction in the form of tidal tails, shells and streams.}
 \label{fig:HCG88}}
\end{figure}

Obviously, the C28IL could be very useful is discovering low surface brightness extended features around galaxies, as well as for other studies. This potential is demonstrated in Figure~\ref{fig:HCG88} that reveals galaxies in the Hickson Compact Group 88 (HCG88) showing clear signs of interaction in the form of a tidal tail, external shells, etc. These features have not been previously reported and will be the topic of a paper now in preparation. Verdes-Montenegro et al. (2001) proposed that HGC88 should be considered to be at a very early stage of interaction based on HI mapping, but this does not seem to be the case. Other contributions from this telescope, indicating its potential to provide good science, rely on long time series; these data have not yet been fully collected and analyzed.

\section{Lessons learned}
\label{sec.lessons}
In this section we describe difficulties encountered while implementing this observatory, and means we adopted to mitigate them.
\begin{itemize}

\item We purchased the C28 essentially as a kit. The telescope was fully assembled, adjusted and tested on the night sky at the manufacturer's facility in the US, then it was partly dis-assembled, crated, and shipped. We found that the assembly was difficult and that, following the re-assembly, the telescope needed significant adjustments and tunings. These would have been avoided, and significant observing time would have been gained, if we would have insisted that the manufacturer would come to the site for a week or so, to assemble and fine-tune the telescope.

\item If telescope optical alignment is required, the user needs a good-quality alignment tool, such as a laser that can be sited well in the focal plane where it is both firmly held so that lateral motions and tilts are prevented, and where it can also be rotated 360$^{\circ}$ around itself, to check that the laser beam is aligned with the telescope optical axis.

\item The C28IL mirror has a protected aluminum coating. While more efficient and durable than bare aluminum, this is also much more difficult to remove and recoat the mirror, if needed. In fact, no local company is willing to perform this task and if recoating would be required, the mirror would have to be sent back to the US incurring a long telescope downtime. Fortunately, our experience with the C18 showed that only regular mirror washing is necessary, at least for the first ten years of the telescope operation.

    \item The telescope uses phosphoric bronze bands on the aluminum drive disks to provide stiffer RA and DEC motions. The ``first generation'' belts were not as stiff as necessary; the manufacturer mailed replacement belts and instructions how to modify our spare belts to the new design. Even so, we have to replace the belts at yearly intervals.

\item The CCD-to-telescope adapter we required for the FLI camera was significantly different from the one supplied by the manufacturer, which could fit a standard SBIG large-format CCD. This resulted in significant observing time lost as we tried to make an adapter of the proper length to allow good focus over the entire field and for different filters.

    \item Following the advice of the owner of the C28US, as well as that of the manufacturer, we installed an air conditioner that reduces the dome internal temperature during daytime. A fan blows cold air from the a/c to the back of the mirror during daytime, so that at the beginning of the night the mirror is fairly cold and the telescope does not require much time to be ready for observations. The a/c and fan are switched on and off by a timer adjusted to a seasonal schedule.

\item We encountered a number of instances with the ScopeDome software failing or freezing. Each have been dealt with by the company, sometimes using remote diagnosis, and implementing updates and new versions. We found that the direct USB connection to the ScopeDome CS is very sensitive to other USB devices that are connected to adjacent ports. %The CS failed many times when connected directly to the computer.
     Routing this connection via a separate USB card in the computer solved this issue.

\item Many cases of faults were traced to bad quality USB cables that connect various components of the system. Only high-quality cables should be used to avoid these faults.

\item Originally, we planned to sit the telescope ``feet'' right on the concrete slab but later, when we had to repeat the polar alignment, we decided to rest the feet on steel plates that allow smooth sliding of the telescope in azimuth.

\item The assembly of the dome itself was a difficult experience. The dome segments are large and heavy, requiring a crane to lift them, stay ropes to maintain the proper orientation, and a number of persons to simultaneously apply the silicone glue/sealant and tighten the bolts. Altogether, there are more than 1200 bolts of different sizes connecting all the dome parts. In retrospect, it would have been better to have the company send an experienced mechanic to oversee the dome construction efforts.

\item Since the ScopeDome was ordered with an entrance door, and this starts way above the ground level (the concrete slab) we fabricated a set of metal stairs that were fixed on the inside of the dome where the door is with the dome parked. On the outer part of the dome, and at the same position of the dome, we installed an inclined- plane metal ramp. This allows  safe human access without having to ``jump'' into the dome through the partly-opened shutter.

\item We encountered an incident with the ScopeDome shutter dis-engaging from its drive wheel and falling to the back of the dome. Fortunately, nobody was hurt in this incident but we were lucky that we had previously installed the inclined plane access ramp to the back of the dome; this stopped the shutter from falling completely to the ground. The solution was to install additional sensors on the dome to determine whether the shutter was at the top or the bottom of its travel, as well as mechanical stops to prevent the shutter from sliding down more than necessary and to be retained close to the dome at all times.

    \item A very useful addition to the original dome was a contact switch to the entrance door. This prevents the opening of the shutter if the door is not fully closed, since these two hardware pieces can conflict.

\item Originally, we planned to install a UPS to service the dome shutter, to be located on the movable part. At this location the UPS is not supplied from the mains unless the dome is in its parking position. We found that regular, even high-capacity, UPS devices discharged fully during the night due to high internal power consumption of the UPS itself and  by the radio link between the two parts of the CS of ScopeDome. In order to have an assured power supply to the shutter, ``deep discharge'' batteries were recommended. These, however, are quite expensive thus this issue is not yet solved.

    \item Following the advice of a customer who purchased the same dome, we used plenty of silicone glue/sealer when assembling the dome segments. Essentially, every contact surface was liberally coated with silicone glue before bolting it in place, and the sealant was allowed to flow outside the joint when tightening the bolts. This ensured that even during intense rain periods no water entered the dome.

    \item Since the Wise Observatory is located in a desert, we experience sometimes dust storms and the construction of the dome, as supplied, would not prevent large quantities of dust from entering the telescope area. To prevent this, we use plastic brushes wherever there is a potential dust access location. We use brushes with four-inch long hairs affixed to the dome skirt, which are in contact with the concrete slab, and with two-inch hairs fixed to the top and bottom of the shutter.

\end{itemize}

\section{Summary}
\label{sec.summary}

We described here the new 28-inch reflector of the Wise Observatory, designed from the start to be a robotic installation. We listed the various components that make this into a modern observatory, including the dome, CCD camera, and operating software packages. Choosing off-the-shelf components and software packages reduced considerably the cost while providing essentially the same features as much more expensive but custom-made telescopes provide.

\section{Acknowledgements}
We are grateful for the financial support toward the C28IL telescope received from Mike Shara, Ezra Drucker, Mike Rich and from the Weizmann Institute of Science. The various stages of assembling the dome and installing the telescope could not have been completed without the invaluable help of Ezra Mash'al and Sammy Ben Guigui from the observatory staff and of Assaf Berwald. We gratefully acknowledge the Mizpe Ramon municipality for loaning the crane that was used to assemble to dome and to bring in the heavy parts of the telescope.%, as well as its operator Mr.?? ??.

%{\bf Shai's additional change: In Fig. 3 mention the STi as "Guider STi"}

\begin{description}{}

\item Akerlof C.~W., et al., 2003, PASP, 115, 132

\item
Bakos G., Noyes R.~W., Kov{\'a}cs G., Stanek K.~Z., Sasselov D.~D., Domsa
I., 2004, PASP, 116, 266

\item Brosch
N., 1992, QJRAS, 33, 27

\item Brosch, N., Polishook, D., Shporer, A., et al.\ 2008, \apss, 314, 163

\item Gregg M.~D., West M.~J., 1998, Natur, 396, 549

\item Landolt A.~U., 1983, AJ, 88, 439

\item Pepper J., et al., 2007, PASP, 119, 923

\item Pollacco D.~L., et al., 2006, PASP, 118,
1407

\item Rich R.~M., Collins M.~L.~M., Black C.~M., Longstaff F.~A., Koch A., Benson A., Reitzel D.~B., 2012, Natur, 482, 192

\item Steele I.~A., et al., 2004, SPIE, 5489, 679

\item Tello, J.C., Riva, A., Hiriart, D., Castro-Tirado, A.J., 2014, Rev. Maxicana Astron. Astrophys., 45 (December 2014)

\item Verdes-Montenegro L., Yun M.~S., Williams B.~A., Huchtmeier W.~K., Del Olmo A., Perea J., 2001, A\&A, 377, 812

\item Yost S.~A., et al., 2006, AIPC, 836, 349

\item Zerbi R.~M., et al., 2001, AN, 322, 275

\end{description}

\end{document}